\documentclass[aps,prl,superscriptaddress,showpacs,twocolumn]{revtex4}
\usepackage{epsfig}
\begin{document}            
\title{Percolation model for a selective response of the resistance of composite semiconducting np-systems towards reducing gases}
\author{Stefanie Russ} 
\affiliation{Institut f\"ur Theoretische Physik,  Arnimallee 14, Freie Universit\"at Berlin, 14195 Berlin, Germany}
\date{\today}

\begin{abstract}

It is shown that a two-component percolation model on a simple cubic lattice can explain an experimentally observed behavior \cite{Savage}, namely that a network built up by a mixture of sintered nanocrystalline semiconducting n- and p-grains can exhibit selective behavior, i.e. respond with a resistance increase when exposed to a reducing gas A and with a resistance decrease in response to another reducing gas B. To this end, a simple model is developed, where the n- and p-grains are simulated by overlapping spheres, based on realistic assumptions about the gas reactions on the grain surfaces. The resistance is calculated by random walk simulations with nn-, pp- and np-bonds between the grains and the results are found in very good agreement with the experiments. Contrary to former assumptions, the np-bonds are crucial to obtain this accordance.

\end{abstract}

\maketitle

\section{I. Introduction}

The percolation model is well-known for its ability to describe a phase transition between an insulating and a conducting phase and has already been used to describe insulator-conductor transitions in electrolytes, two-component conductors, binary glasses and more (for an overview see e.g. \cite{BH}). In \cite{Ulrich}, it was proposed that a disordered system of sintered nanocrystalline semiconducting n-grains should act as a gas sensor \cite{Heiland,Kohl01,Morrison87,Tiemann}, i.e. switch from the insulating to the conducting phase in response to a reducing gas. The mechanism based on the percolation effect is the following: In normal air, according to a standard model \cite{Park,Ulrich}, oxygen adsorbs at the grain surface and traps mobile electrons (see Fig.~\ref{bi:grains}(a)), so that smaller grains are insulating. When the oxygen coverage is reduced by a reaction of the oxygene with a reducing gas, more and more grains become conducting until the concentrations of conducting grains and bonds both exceed the so-called critical concentrations \cite{Draeger}. Then, the resistance $R$ of the system jumps to a much smaller order of magnitude allowing for the detection of the gas. (For an overview of percolation effects in gas sensors see \cite{SauRu}).

However, this transition alone delivers no information on the type of the reducing gas. The ability of a sensor to answer differently to different gases is named ''selectivity'' and is -- even if highly desirable  -- difficult to achieve \cite{Mandayo03,Heilig}. One approach is to compose a sensor of two types of grains, one of them decreasing and the other one increasing $R$ with increasing gas concentration, as e.g. semiconducting n- and p-grains \cite{Choi,Savage}. Such a composite system should be able to respond with a resistance increase as well as with a decrease to the gas exposure and even a mutual neutralisation of both effects is conceivable. The challange is to find one specific system that can show both, increase and decrease of $R$, depending on the gas.
Experimentally, such a behavior has been reported on systems of sintered $\mbox{TiO}_2$ grains in the two representations anatase and rutile \cite{Savage} that have been reported as n- and p-semiconductors, respectively. 
A sample with a volume ratio 1:3 of n-anatase to p-rutile was found to show (nearly) no response to $CH_4$ but react with a resistance decrease to $CO$. Percolation effects have been assumed responsible for this interesting behavior.

\unitlength 1.85mm
\vspace*{0mm}
\begin{figure}
\begin{picture}(40,15)
\def\epsfsize#1#2{0.3#1}
\put(0,0){\epsfbox{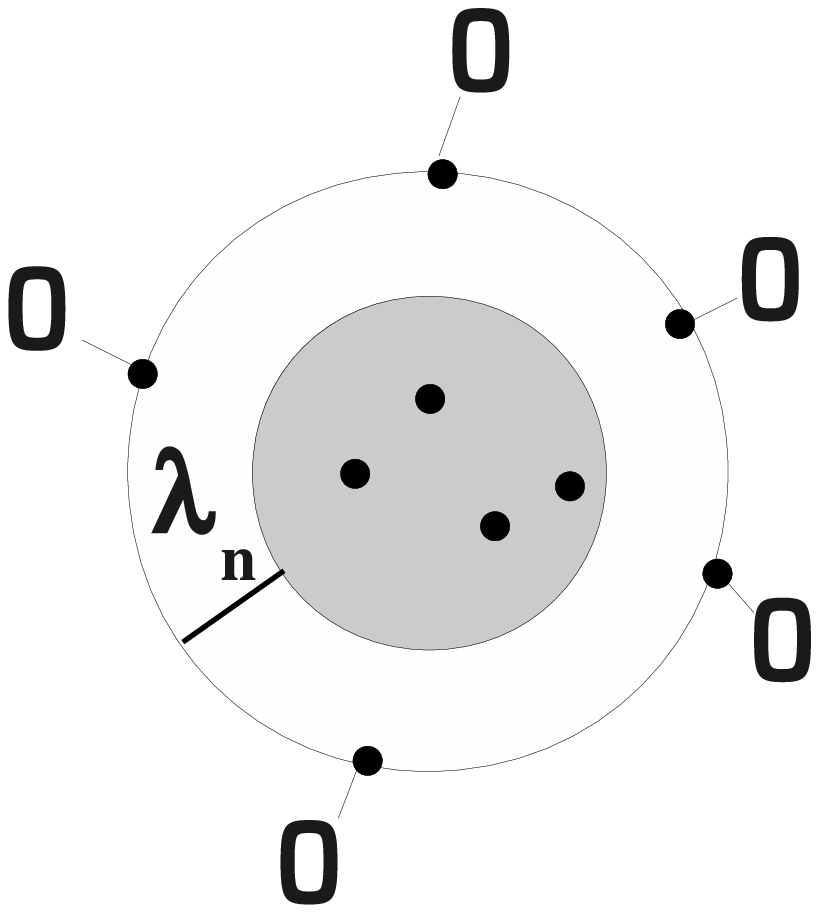}}
\put(0,0){\makebox(1,1){\bf\large (a)}} 
\put(20,0){\epsfbox{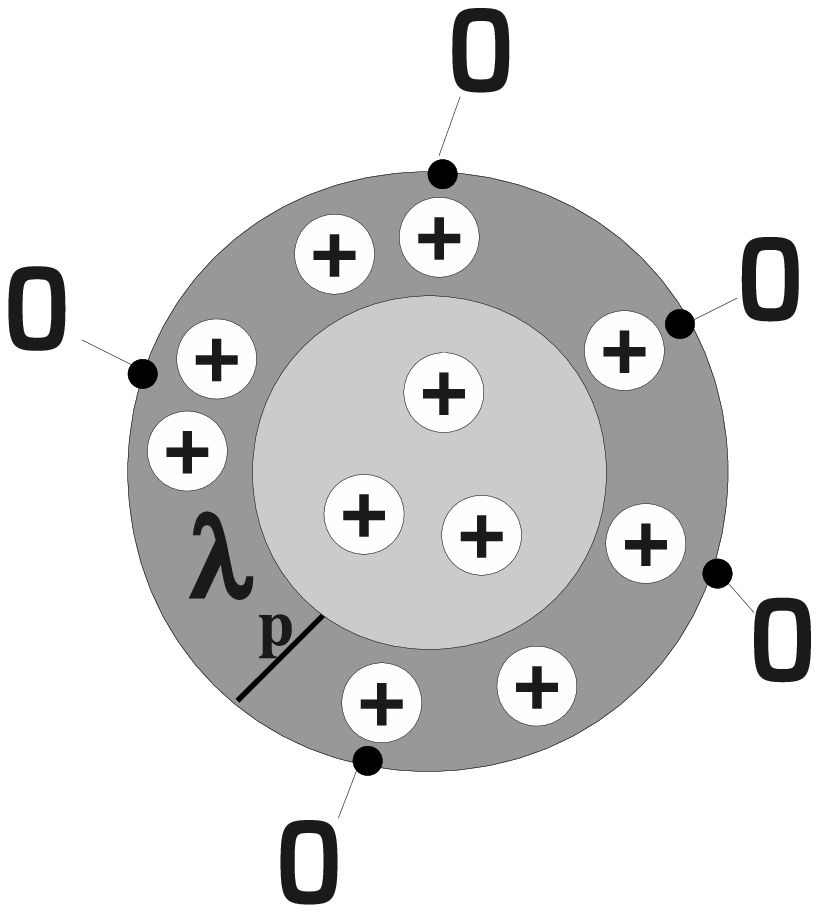}}
\put(21,0){\makebox(1,1){\bf\large (b)}} 
\end{picture}
\caption[]{\small (a,b): Schematized models of (a) an n-grain and (b) a p-grain, covered by adsorbed oxygen that traps electrons on the surface (small black dots), thereby becoming ionized and creating depletion-/enrichment shells, respectively. The depletion shell of (a) with thickness $\lambda_n(0)$ contains no electrons, whereas the enrichment shell of (b) with thickness $\lambda_p(0)$ has an increased hole density as compared to the bulk. 
\label{bi:grains}}
\end{figure}

In this paper, I develop a simple model, based on the percolation model, that captures the essential properties of composite gas sensors and show by computer simulations that different coexisting percolating pathways can indeed cause a selective behavior towards different reducing gases. Similar to the observations of \cite{Savage} I show that certain composite systems can show no reaction or even an increase of $R$ to a gas A and a decrease of $R$ to another gas B. However, contrary to the assumptions of \cite{Savage}, I show that bonds between n- and p-grains play a major role for a selective behavior as they couple the two types of pathes.
Both types of grains are modelled in reasonable approximations under the influence of adsorbed oxygen, thereby applying basic microscopical concepts \cite{Lee,Morrison_Buch,Park,Barsan}. Different gases are distinguished by their reaction rates with the adsorbed oxygen, expressed by appropriate ''reaction factors'' $\alpha_n$ and $\alpha_p$ that describe the ratio of oxygen that reacts with the considered gas on n- and on p-surfaces, respectively. All grains are now conducting (in contrast to the works of \cite{Ulrich,Draeger}), i.e. the n-grains are not too small. 

The paper is organized as follows: In section II, the microscopical picture of the single grains is explained, while section III describes how the grains are organized in networks and how the resistances are computed. Section IV presents the results and in the final section V, the specifics and simplifications of this model are discussed, as well as possible extensions.

\section{II. Microscopical picture}

\subsection{Surface effects on n-grains}

The standard Schottky model of an n-grain under the influence of oxygen \cite{Kohl89,Kohl01,Tiemann,Kolmakov09,Park,Draeger} at not too high temperatures (where surface- are dominant over bulk reactions) is depicted in Fig.~\ref{bi:grains}(a): in normal air, oxygen is adsorbed on the grain surface and traps electrons from the conduction band, so that the grains are covered by oxygen ions. This way, a negative surface charge arises at the grain surfaces that is accompanied by a positively charged depletion layer of thickness $\lambda_n(0)$ (space charge layer), where the argument $0$ stands for zero concentration of the reducing gas. The inner core of the grains is considered as unchanged with an electron density that equals the volume density, $N_D$, of donor atoms that are considered as fully ionized. In Schottky approximation, one assumes that the density $N_D^-$ of mobile electrons in the depletion layer is zero and finds $\lambda_n(0)=(2\epsilon_0\epsilon_reV_s/N_De^2)^{1/2}$ by solving the 1D-Poisson equation, where the work function $eV_s$ between the surface and the core is of the order of $1\,eV$ or somehow below \cite{Morrison_Buch}. Here, we will rely to the values of $\lambda_n(0)\approx 10\,nm$ and $N_D=3.5\cdot 10^{-3}\,nm^{-3}$ that have already been used in \cite{Ulrich,Kohl01,Draeger,Wang,Yamazoe91}. From $\lambda_n(0)$ and $N_D$, one gets the surface density of the (initially) adsorbed oxygen by the condition of ''charge neutrality'' as $N_{\rm{ox,n}}^{0}\approx \lambda_n(0) N_D = 3.5\cdot 10^{-2} nm^{-2}$.

When a reducing gas as e.g. $CO$ is offered, according to a standard model \cite{Park,Ulrich}, it reacts with part of the adsorbed oxygen according to $O^-+CO\to CO_2+e^-$ and releases the electrons back to the bulk, so that the negative surface charge is reduced. Now, the charge neutrality delivers $\lambda_n(N_r)$ as a function of the surface density $N_r$ of the reducing gas,
\begin{equation}\label{lambdan}
\lambda_n(N_r)\approx \frac{N_{\rm{ox,n}}^{0}-\alpha_nN_r}{N_D} = \lambda_n(0)\cdot(1-\frac{\alpha_n N_r}{N_{\rm{ox,n}}^{0}}).
\end{equation}
The reaction factor $\alpha_n$ is related to the equilibrium rate of the reaction between oxygen and reducing gas and describes the fraction of gas that reacts with the adsorbed oxygen. A simpler version of Eq.~(\ref{lambdan}) has already been used in \cite{Ulrich,Draeger} with $\alpha_n=1$, i.e. under the assumption that each gas molecule reacts with one oxygen atom. As surface reactions on different types of grains are compared here, I apply the more realistic assumption that $\alpha_n\le 1$. Clearly, $\alpha_n$ differs from the respective reaction factor on p-grains. 

\subsection{Surface effects on p-grains}

Also the p-grains can be described by a positively charged shell and a neutral core \cite{Barsan} (see Fig.~\ref{bi:grains}(b)), where the volume density of positive holes in the core equals the acceptor density $N_A$. 
When oxygen adsorbs at the surface of p-grains, it traps electrons from the valence band thereby creating additional holes. Therefore, also on the surface of p-grains a positively charged shell arises but instead of a depletion layer (as on n-grains) it represents an accumulation layer. This means that the density of mobile holes $N_A^+$ inside the p-shell is even increased as compared to the hole density of the core, i.e. $N_A^+>N_A$. $N_A^+$ depends on the potential $V_s$ and the temperature $T$ via the Boltzmann distribution. Contrary to the value of $N_D^-$ (on n-grains) that is simply set to zero, the mean density $N_A^+$ of mobile holes in the accumulation layer must also be determined. This has been done in \cite{Barsan} based on prior considerations from \cite{Morrison_Buch} under the assumption that the grain diameter $D$ is much larger than the thickness of the accumulation layer, yielding 
\begin{equation}\label{p_grains}
N_A^+ = N_A\left[\left(\exp(\frac{eV_s}{k_BT}) - \frac{eV_s}{k_BT} -1\right)^{1/2} + 1\right].
\end{equation}
(In the following, I account for the condition $D\gg\lambda_p$ by choosing $\langle D\rangle\ge 100\,nm$.)
Also other parameters for p-grains can be taken from \cite{Barsan}: The thickness $\lambda_p(0)$ of the accumulation layer in normal air is normally attributed to the Debye length $L_D=(\epsilon_0\epsilon_rk_BT/N_Ae^2)^{1/2}$ with the Boltzmann constant $k_B$ as $\lambda_p(0)\approx\sqrt{2}L_D$, while $eV_S$ has been specified as between $0.3$ and $0.6\,eV$ and $N_A$ between $1.4\cdot 10^{-4}\,nm^{-3}$ and $10^{-1}\,nm^{-3}$. With $T\approx 500\,K$, this leads to $\lambda_p(0)$ between $0.7\,nm$ and $16.5\,nm$ and a ratio $N_A^+/N_A$ between $30$ and $1000$. 
The surface density of initially adsorbed oxygen is again obtained via the charge neutrality as $N_{\rm{ox,p}}^{0}=\lambda_p(0)\cdot(N_A^+-N_A)$, leading to values of $N_{\rm{ox,p}}^{0}$ between about $8\cdot 10^{-2}\,nm^{-2}$ and $70\,nm^{-2}$, which means that $N_{\rm{ox,p}}^{0}\gg N_{\rm{ox,n}}^{0}$, as already pointed out in \cite{Kohl01}. In order to control the response of the systems by $\lambda_p$, I rely to those values (inside this range), where $\lambda_p(0)$ is not too small, i.e. $N_A$ not too large.

Under the influence of a reducing gas, $\lambda_p(N_r)$ can be estimated in complete analogy to (\ref{lambdan}) as
\begin{equation}\label{lambdap}
\lambda_p(N_r)\approx \frac{N_{\rm{ox,p}}^{0}-\alpha_pN_r}{N_A^+-N_A} =
\lambda_p(0)\cdot(1-\frac{\alpha_p N_r}{N_{\rm{ox,p}}^{0}}),
\end{equation}
where $\alpha_p$ describes the reaction rate between the reducing gas and the oxygen on the surface of p-grains.
We can see from Eqs.~(\ref{lambdan}) and (\ref{lambdap}) that the thicknesses of the shells are largest, when the grain surfaces are covered solely by adsorbed oxygen, i.e. when $N_r=0$.

\section{III. Numerical Simulations}

\subsection{The random walk method} 

The mapping of a complicated composite system onto a percolation problem has been discussed for several decades \cite{Berlyand,Feng,ChenSchuh,Sihvola,Helsing,Willot}. In a general system of overlapping conducting particles inside a matrix of lower conductance, different types of contacts between particles may occur. A model example  is the ''checkerboard'', i.e. a 2D or 3D system of squares or cubes of different materials that touch each other by nearest-neighbor contacts (via the borders) and by next-nearest neighbor point contacts via the corners (where pathes around the edges through neighboring units may be involved \cite{Feng}). It has been shown since long that percolation thresholds and exponents of these systems differ from the ones of the corresponding properties of percolating networks, where the squares (or cubes) have been replaced by discrete sites connected by bonds and that the corner contacts between next-nearest neighbors may even lead to a second percolation threshold \cite{Berlyand,Feng,ChenSchuh}. Different ''mixing rules'' telling us, in which geometrical shapes and with which connections disordered heterogeneous media should be mapped onto lattices have been established \cite{Sihvola} and tested by comparing the numerical solutions of different physical properties, as e.g. conductance or admittance gained on discretized lattices to the exact solutions \cite{Helsing,ChenSchuh}. However, while checkerboard problems can be solved analytically and numerically by discretizing the systems \cite{Helsing}, for the majority of systems, clear mixing rules are not available \cite{Sihvola}.

In the systems of this work, n- and p-grains with similar conductances are sintered together. We consider well-sintered grains, where the electron hopping across grain-boundaries can be neglected \cite{Wang,Yamazoe91}, especially when the grain surfaces are rough \cite{sap96}. 
Conductance over next-nearest neigbors by a direct (small) overlap or a point contact is in principle possible if both next-nearest neighbors are big enough, but for the system parameters chosen here the contribution to the conductance can be considered as much smaller than conductance over nearest neighbors (see below for the sintering parameter $\Theta$). 

Here, it is not our aim to reproduce the exact resistance of a given system, but to understand the qualitative behavior, namely if percolation effects are able to influence the reaction of such systems towards {\it different gases} in a way that the {\it same} system can show  increasing, decreasing or constant resistance as a function of the surface gas density.
Therefore, we consider the relative resistance $R(N_r)/R(0)$, i.e. the ratio between the resistances with and without gas, so that all specific material properties that vary only slowly with $N_r$ cancel out. The model system has been maintained simple: the resistances have been calculated by random walk simulations \cite{BH,gefen,SauRu} on a disordered cubic lattice, where the sites represent simplified n- or p-grains, bonds between nearest-neighbor sites have been estimated from the area of the conducting overlap between them and next-nearest neighbor contacts have been neglected. To this end, the nanogranular films are mapped onto a simple cubic network of size $N\times N\times N_z$ and lattice constant $a$, consisting of sites and bonds between the sites. $N_z$ is the number of monolayers and $N\gg N_z$. 
Random walk simulations are a well-explored method to calculate the specific resistance $\rho$ of discretized disordered systems. I shortly illuminate the random walk method with the help of Fig.~\ref{bi:rw}, where a single random walk of $50$ time steps is shown on a square lattice. The 1st 50 steps of this walk are shown in Fig.~\ref{bi:rw}(a) in a schematized picture, while \ref{bi:rw}(b) shows $r(t)\equiv \sqrt{r^2(t)}$ of this walk as a function of $t$. The walker starts at the lattice site labeled with ''0'' and performs an irregular trajectory $\vec r(t)$.  Finally, in (c) we see the ''mean square displacement'' $\langle r^2(t)\rangle$ averaged over $1$, $10^2$ and $10^4$ realizations and can see that in average $\langle r^2(t)\rangle$ gives a straight line, i.e. $\langle r^2(t)\rangle\sim t$.

\unitlength 1.85mm
\vspace*{0mm}
\begin{figure}
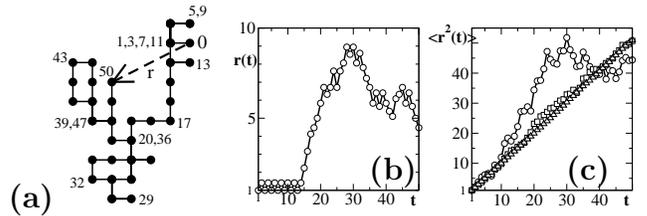

\begin{picture}(40,10)
\def\epsfsize#1#2{0.3#1}
\put(1,0){\epsfbox{Fig2a.eps}}
\put(-1,0){\makebox(1,1){\bf\large (a)}} 
\def\epsfsize#1#2{0.2#1}
\put(14,0){\epsfbox{Fig2b.eps}}
\put(25,2){\makebox(1,1){\bf\large (b)}} 
\put(28,0){\epsfbox{Fig2c.eps}}
\put(39,2){\makebox(1,1){\bf\large (c)}} 
\end{picture}
\caption[]{\small Illustration of the random walk method: (a) Sketch of a 2D-random walk on a square lattice (lattice sites not shown) starting at lattice site labeled ''0'' and ending at site labeled ''50'', where the numbers represent the time steps. At some selected lattice sites, all step numbers are displayed, i.e. ''20,36'' indicated that the walker passes at this site at time step 20 and 36. The vector $\vec r(t=50)$ of length $r$ is indicated by the arrow. (b) The length $r(t)=\sqrt{r^2(t)}$ is shown for the walk of a.) as a function of $t$ for the 1st 50 time steps. (c) $r^2(t)$ is shown as a function of $t$ for the 1st 50 time steps as an average over $10$, $10^2$ and $10^4$ random walks (circles, squares and triangles, respectively). 
\label{bi:rw}}
\end{figure}

The macroscopic resistance $R$ of a layer of cross-section $A$ and length $L$ is proportional to the specific resistance $\rho$ by $R=\rho L/A$, where $\rho$ is inversely proportional to the diffusion coefficient (and thus to the mobility of the charge carriers), via the Einstein relation \cite{Shlomo,Diff_coeff}.
Different definitions of the diffusion coefficient may come into play and the underlying assumption of the random walk method is therefore the equivalence between the so-called ''transport diffusion coefficient'' $D_t$ and the ''self- or tracer-diffusion coefficient'' $D_s$. $D_t$ connects the current density $\vec j$ to the gradient of the particle density $\vec\nabla c$ via $\vec j=-D_t \vec\nabla c$, while $D_s$ connects the mean square displacement $\langle r^2(t)\rangle$ to the diffusion time t via 
\begin{equation}\label{einstein}
\langle r^2(t)\rangle=2dD_st
\end{equation} 
with the Eucledian dimension $d$ (cf. Fig.~\ref{bi:rw}(c)). The equivalence between $D_s$ and $D_t$ has sometimes been questioned, but in the end always found valid for the case of non-interacting particles \cite{russ,comets}. This means that $D_s=D_t \propto 1/\rho$, enabling us to calculate $\rho$ from the diffusion coefficient $D_s$. 
In $TiO_2$ nanoparticle networks, however, an additional effect might come into play, namely the dependence of the electron density on the chemical potential \cite{Diff_coeff}. It has become common to introduce the ''chemical diffusion coefficient'' $D^*$ that differs from $D_t$ by the ''thermodynamic factor'' that accounts for deviations of the electron density from Boltzmann statistics \cite{Diff_coeff}. As the Boltzmann statistics is inherent to Eq.~\ref{p_grains}, we rely to $D_s$ from Eq.~(\ref{einstein}).

\unitlength 1.85mm
\vspace*{0mm}
\begin{figure}
\begin{picture}(40,10)
\def\epsfsize#1#2{0.3#1}
\put(1,0){\epsfbox{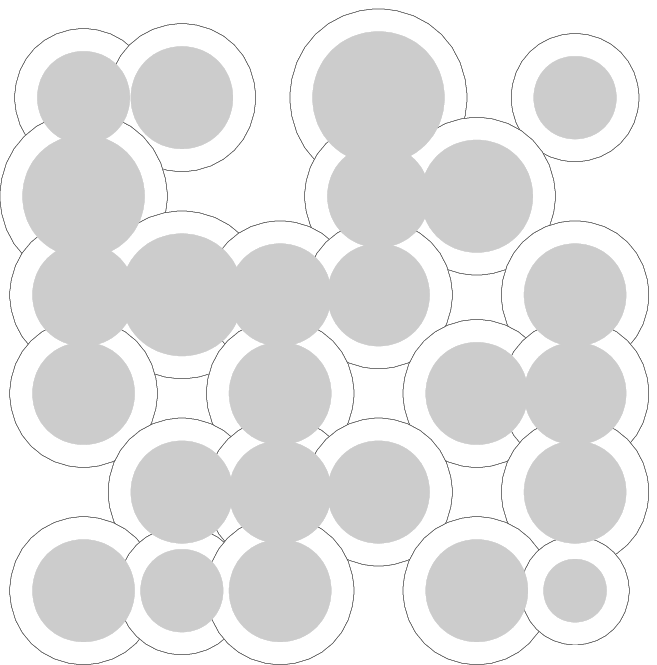}}
\put(-1,5){\makebox(1,1){\bf\large (a)}} 
\put(17,0){\epsfbox{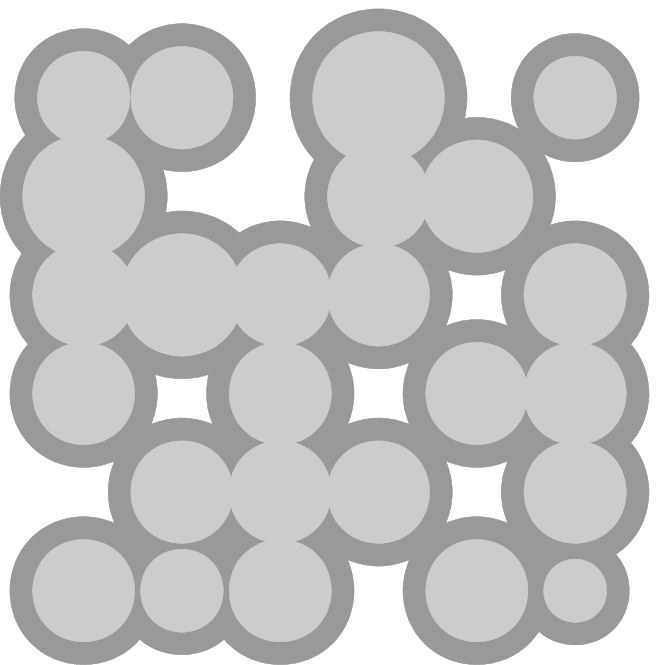}}
\put(15,5){\makebox(1,1){\bf\large (b)}} 
\put(32,0){\epsfbox{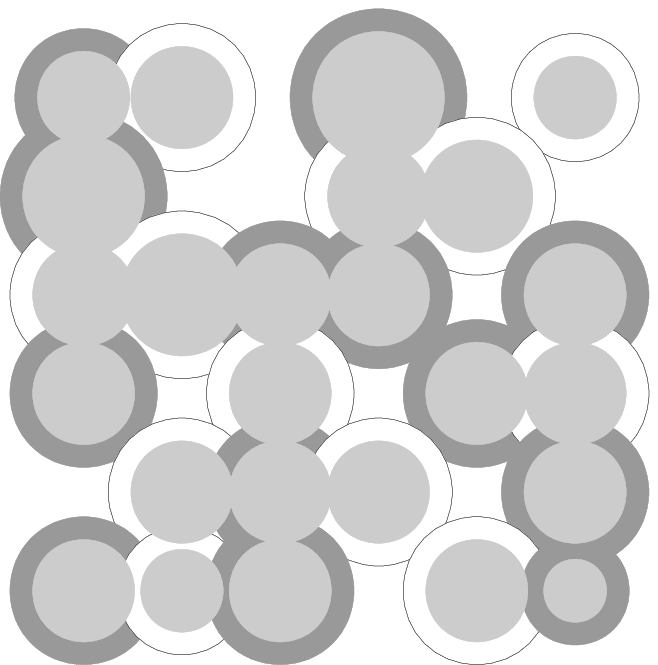}}
\put(30,5){\makebox(1,1){\bf\large (c)}} 
\end{picture}
\caption[]{\small 2D-sketches of networks of (a) n-grains (grey core with white depletion shell), (b) p-grains (grey core with dark grey enrichment shell) and (c) heterogeneous networks of both type of grains.
\label{bi:networks}}
\end{figure}

\unitlength 1.85mm
\vspace*{0mm}
\begin{figure}
\begin{picture}(40,5)
\def\epsfsize#1#2{0.18#1}
\put(0,0){\epsfbox{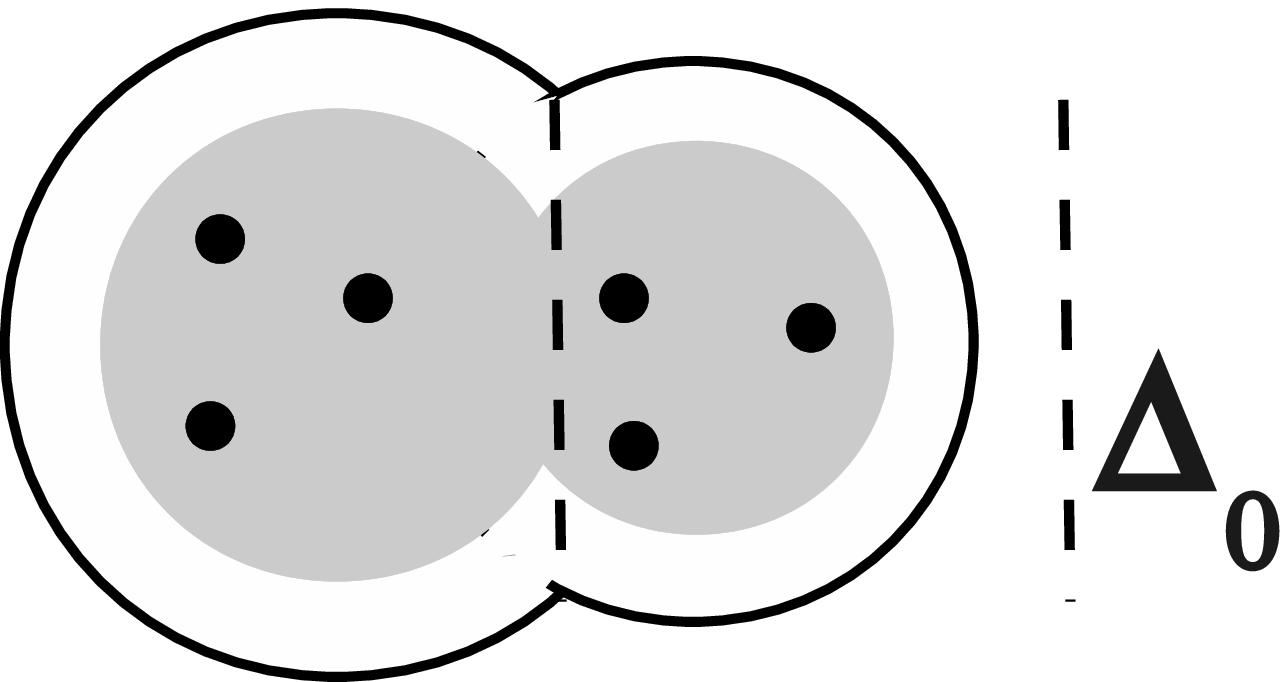}}
\put(16,0){\epsfbox{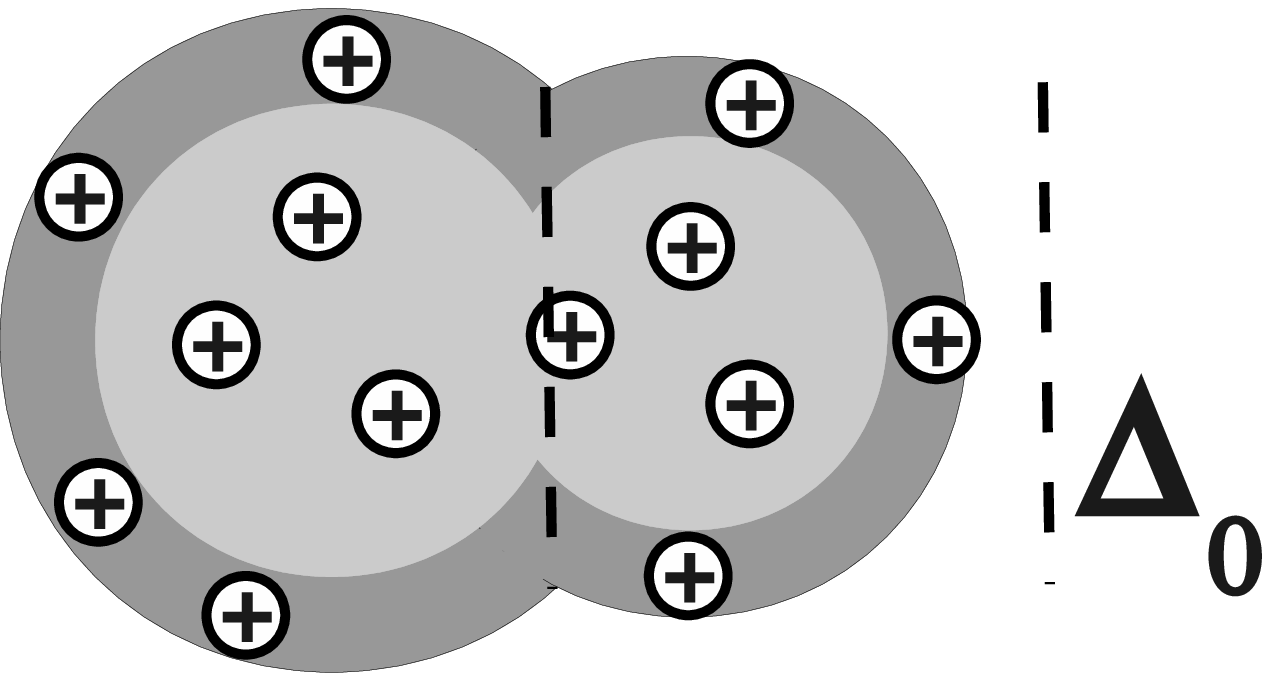}}
\put(31,0){\epsfbox{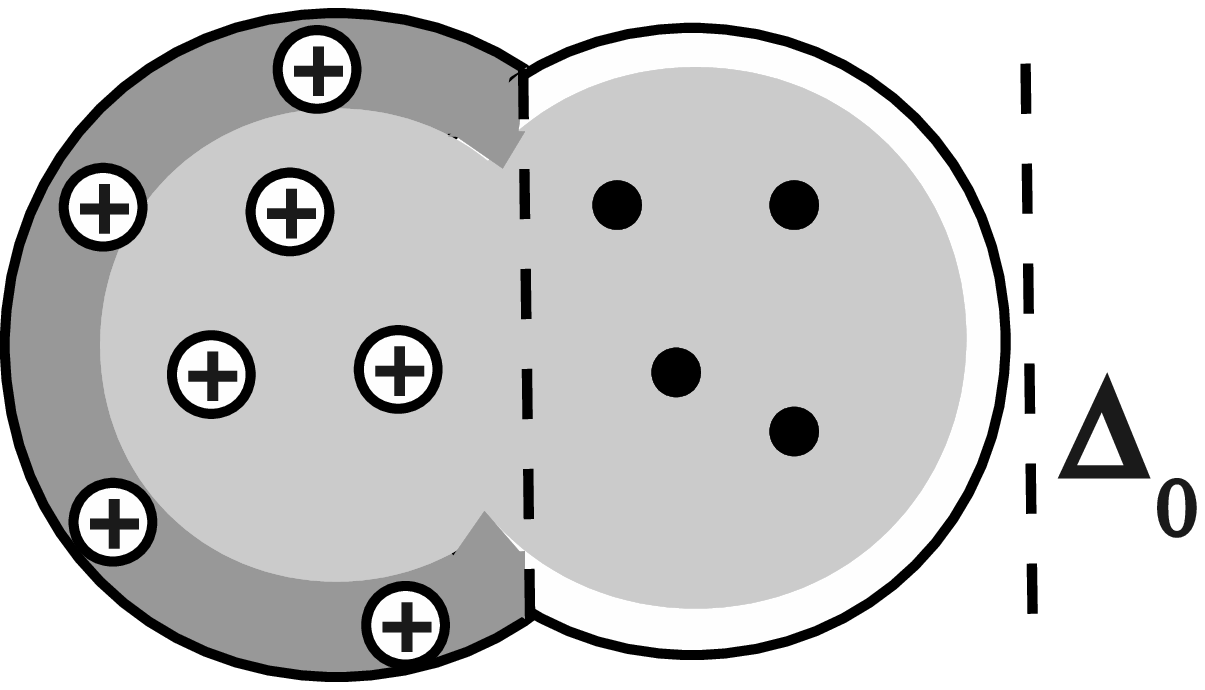}}
\put(-2,0){\makebox(1,1){\bf\large (a)}} 
\put(14,0){\makebox(1,1){\bf\large (b)}} 
\put(30,0){\makebox(1,1){\bf\large (c)}} 
\end{picture}
\caption[]{\small Schematized models of overlapping grains with a geometrical neck of diameter $\Delta_0$ between them: (a) nn-bond, (b) pp-bond and (c) np-bond.
\label{bi:bonds}}
\end{figure}

Similar to \cite{Draeger}, the lattice points are occupied by spherical grains (with probability $p$), or left empty (with probability $1-p$), where $p\ge 0.8$ in the following. Schematic pictures of these networks are shown in Fig.~\ref{bi:networks}.
The diameters $D_i$ of the individual grains $i$ are chosen randomly from a homogeneous distribution of width $w=0.6\,a$ around a mean diameter $\langle D\rangle$, so that neighboring spheres can overlap, thereby forming connections in the shape of necks with a circular cross-section of (geometrical) diameter $\Delta_0$ between them (see Fig.~\ref{bi:bonds}).
The mean values of grain and neck diameters are related via the sintering parameter $\Theta$ as $\langle \Delta_0\rangle\simeq \Theta \,\langle D\rangle$ that is chosen as $\Theta=1/\sqrt{2}$, close to experimentally known values \cite{Ulrich}. With this value of $\Theta$, next-nearest neighbors of a monosized lattice would not overlap but just touch each other.

The grains are randomly divided into n-grains with probability $p_n\in\{1,0.75,0.5,0.25,0\}$ and p-grains with probability $p_p=1-p_n$, which corresponds to the values of \cite{Savage}, where mass ratios of n-to-p grains of $1:0$, $3:1$, $1:1$, $1:3$ and $0:1$ have been considered. 
Assuming a constant number of charge carriers and a constant mobility through the grains, we find $R(N_r)/R(0)$ from the inverse ratio of the diffusion constants $D_s(0)/D_s(N_r)$ that are computed with an average over $10^5$ systems. For this first analysis, differences between hole and electron mobility are neglected as well as the different sizes of n- and p-grains from the experiments of Ref.~\cite{Savage}.

\subsection{Calculation of the jump probabilities}
 
Before performing random walk simulations, we have to define the jump probabilities $p_{ij}$ proportional to the conductances along the bonds, i.e. for nn-, pp- and np-connections between the grains (which we address as $p_{i,j}^{nn}$, $p_{i,j}^{pp}$ and $p_{i,j}^{np}$, respectively). 
Fig.~\ref{bi:bonds} shows how the bonds between the grains are modelled microscopically by following the lines of Refs.~\cite{Draeger,Lee}. 
In the case of two n-grains -- normally of unequal sizes -- current can only flow through the conducting core-core overlap (channel) with the reduced diameter $\Delta_{nn}=\Delta_0-2\lambda_n(N_r)$. The jump propability $p_{i,j}^{nn}$ therefore scales with the density $N_D$ of charge carriers and the area $\Delta_{nn}^2\pi/4$, leading to $p_{i,j}^{nn} = C_n \Delta_{nn}^2\pi N_D/4$. The prefactor $C_n$ is introduced for normalization, i.e. in order to keep the sum of the jump probabilities of a given site to all its neighbors well below unity. It is the same for all connections and independent of $N_r$, i.e. it cancels out for $R(N_r)/R(0)$.
The current between neighboring p-grains flows over two parallel paths \cite{Lee}: one through the cores (with a probability that scales with $N_A$ and the area $A_{\rm{ch}}=\Delta_{pp}^2\pi/4$ of the channel) and one over the shells (scaling with $N_A^+$ and the area of the shell-shell-overlap $A_{\rm{sh}}=\left(\Delta_0^2-\Delta_{pp}^2\right)\pi /4$, where $\Delta_{pp} = \Delta_0-2\lambda_p(N_r)$. The jump propability between the p-grains is the sum of both probabilities, i.e. $p_{i,j}^{pp} = C_n \left(A_{\rm{ch}} N_A + A_{\rm{sh}} N_A^+\right)$.
In \cite{Savage}, the np-bonds have been assumed as cut, because of the high resistance of reverse-biased np-contacts. However, as rectifier properties disappear at very small voltages \cite{Dieter_privat}, I consider the np-bonds as conducting and model their jump-probabilities in a simplified way as proportional to the overlap area with diameter $\Delta_{np}$ between the n-core and the total p-grain and to $N_D$, i.e. $p_{i,j}^{np} = C_n \Delta_{np}^2\pi N_D/4$.
Here, the shell of the n-grains is again considered insulating, while the total p-grain is considered conducting, so that in most cases the n-core is the limiting quantity of these bonds (except some cases, where the total p-grains are smaller than the core of the n-grain). To test the influence of the np-bonds, the results are in some cases compared to the case where all np-bonds are cut.

\section{IV. Results}

We now discuss the results gained for not too high gas concentrations $N_r$, i.e. the simulations are stopped when the thickness $\lambda_n(N_r)$ of the depletion layer of the n-grains becomes $0$. For still higher gas concentrations, $\lambda_n$ would no longer play a role for controling the sensor response and $R$ would simply increase with $N_r$, thereby only reflecting the response of the p-grains to the gas.

In all figures, the symbols and colors symbolize the different values of $p_n$, i.e. they stand for systems (from bottom to top) of only n-grains, 1:3 ratio of p:n, 1:1 ratio of p:n, 3:1 ratio of p:n, only p-grains. For the n-grains, I refer to the parameters of \cite{Wang,Yamazoe91} that have also been used in \cite{Ulrich,Draeger} and for the p-grains, I test three parameter sets, close to the ones reported in Ref.~\cite{Barsan} (see above) covering the situations $\lambda_p(0)>\lambda_n(0)$, $\lambda_p(0)\approx\lambda_n(0)$ and $\lambda_p(0)<\lambda_n(0)$. For clarity, $\lambda_n(N_r)$ and $\lambda_p(N_r)$ are shown in the corresponding inset of each figure. In the simulations, I tested many values for $\langle D\rangle\in [50\,nm,200\,nm]$, $p\in [0.6,1]$ as well as variations of $N_z$, $N_D$ and $N_A$. I found that $R(0)$ depends on the specific parameter choice, but the normalized curves $R(N_r)/R(0)$ are qualitatively very similar for all parameters (except for smaller values of $p$, where infinite percolation clusters do not exist in every system). The results shown here should therefore be understood as typical examples.

\subsection{Unity reaction factors}

\unitlength 1.85mm
\vspace*{0mm}
\begin{figure}
\begin{picture}(40,30)
\def\epsfsize#1#2{0.3#1}
\put(0,1){\epsfbox{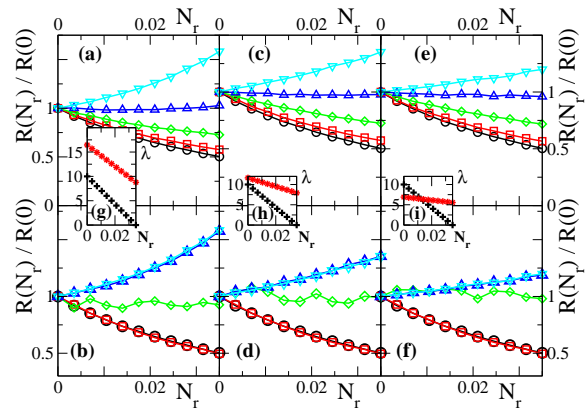}}
\end{picture}
\caption[]{\small [Color online] Under the reaction factors $\alpha_n=\alpha_p=1$, $R(N_r)/R(0)$ is plotted versus $N_r$ for $N_z=9$, $\langle D\rangle=100\,nm$, $T=500K$, $p=0.8$ and $\epsilon=8$. For n-grains we have $N_D=3.5\cdot 10^{-3}\,nm^{-3}$ and $\lambda_n(0)=10\,nm$, while for p-grains, we chose  $qV_s=0.3\,eV$ and different values of $N_A$: (a,b) $N_A=1.4\cdot 10^{-4}nm^{-3}$, (c,d) $N_A=2.9\cdot 10^{-4}nm^{-3}$  and (e,f) $N_A=8\cdot 10^{-4}nm^{-3}$ refering to $\lambda_p(0)>\lambda_n(0)$, $\lambda_p(0)=\lambda_n(0)$ and $\lambda_p(0)<\lambda_n(0)$, respectively. The different colors and symbols indicate (from bottom to top) $p_n=1$, i.e. only n-sites (black circles), $p_n=0.75$ (red squares), $p_n=0.5$ (green diamonds), $p_n=0.25$ (blue triangles up) and $p_n=0$, i.e. only p-sites (cyan triangles down). 
While in the top row (a,c,e), the np-bonds have been included into the calculations, they have been set to zero in the bottom row (b,d,f). 
The insets (g,h,i) show the corresponding values for $\lambda_n(N_r)$ (black $+$) und $\lambda_p(N_r)$ (red stars). 
\label{bi:prepa}}
\end{figure}

\unitlength 1.85mm
\vspace*{0mm}
\begin{figure}
\begin{picture}(40,30)
\def\epsfsize#1#2{0.3#1}
\put(0,1){\epsfbox{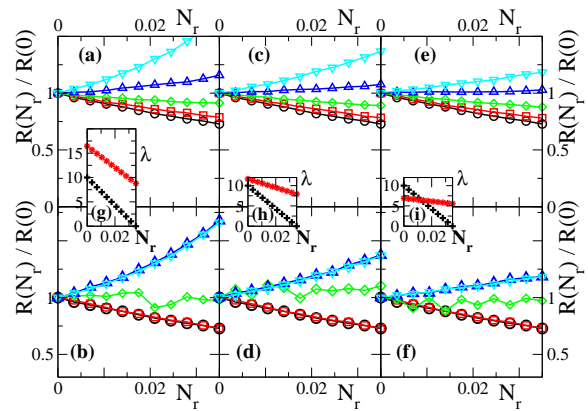}}
\end{picture}
\caption[]{\small [Color online] Under the reaction factors $\alpha_n=\alpha_p=1$, $R(N_r)/R(0)$ is plotted versus $N_r$ for $\langle D\rangle=200\,nm$. All other parameters are the same as in Fig.~\ref{bi:prepa}, as well as the meaning of the different colors, symbols and insets (g-i). 
\label{bi:prepa2}}
\end{figure}
 
Figures~\ref{bi:prepa} and \ref{bi:prepa2} show $R(N_r)/R(0)$ as a function of $N_r$ for systems of mean grain diameters $\langle D\rangle = 100\,nm$ and $\langle D\rangle = 200\,nm$, respectively. For this first test, the reaction factors are set to  $\alpha_n=\alpha_p=1$, i.e. each gas molecule at the (n- or p-) grain surfaces is considered to react with one oxygen ion.  The three columns show the different parameter sets for the p-grains, whereas the n-grains are always the same.
For the simulations shown in the top rows (a,c,e), also the np-bonds have been considered conducting, while they have been considered as cut in (b,d,f) (bottom rows). In the 1st case, in agreement with the experimental data from \cite{Savage}, all curves are seperated well from each other. 
In the second case, the two lowest curves, $p_n\ge 0.75$, (black circles and red squares) of each figure as well as the two top curves, $p_n\le 0.25$, (blue triangles up and cyan triangles down) fall together. With a critical percolation threshold of $p_c\approx 0.3$ it is clear that only the n-paths percolate  for $p_n\ge 0.75$, while only the p-paths percolate for $p_n\le 0.25$. Without np-bonds,  both paths decouple and $R(N_r)/R(0)$ reflects only the change of the dominant grain type with the gas, while the behavior of the non-percolating grain type plays no role. Otherwise speaking, $p_n$ is unimportant and the curves of each type fall together.
A third class of curves (green diamonds) is formed by the systems with $p_n=0.5$, where both paths percolate simultaneously and $R(N_r)/R(0)$ shows an intermediate behavior. However, the range of $p_n$-values with percolating n- {\it and} p-pathes is quite small, given a total occupation probability of $0.8$. 
In summary, without including the np-bonds into the simulations, the experimental results that show individual characteristic curves for each system cannot be found and only three groups of curves appear instead. 

\subsection{Reactions to different gases}

Before discussing the numerical results for different reaction factors, we first take a look on some of the experimental data from Ref.~\cite{Savage} that are redrawn in Fig.~\ref{bi:sav_exp}. Figure~\ref{bi:sav_exp}(a) shows five samples with different mass relations of n-anatase to p-rutile under the influence of one specific reducing gas ($CH_4$). One can see that -- depending on the n-to-p ratio of the sample -- the relative resistance $R/R_0$ may increase or decrease in complete analogy to the numerical data of Figs.~\ref{bi:prepa} and \ref{bi:prepa2}. (The succession of the data with the 3:1 and the 1:1 ratio of n-anatase:p-rutile is interchanged as compared to the expectations, which is commented in \cite{Savage}.) In Fig.~\ref{bi:sav_exp}(b), $R/R_0$ for one of these samples (second from above) is shown again, but this time it is compared to the reaction of the same sample to a second reducing gas ($CO$). One can easily recognize that $R/R_0$ stays (nearly) constant when exposed to $CH_4$, while it decreases drastically when exposed to $CO$.

\unitlength 1.85mm
\vspace*{0mm}
\begin{figure}
\begin{picture}(40,16)
\def\epsfsize#1#2{0.28#1}
\put(0,1){\epsfbox{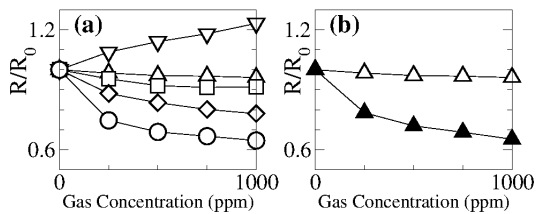}}
\end{picture}
\caption[]{\small Some data of Fig.~5 of Ref.~\cite{Savage} are redrawn: (a) $R/R_0$ under the influence of $CH_4$ for five films of anatase-rutile (n-p) composites with different mass ratios n-anatase:p-rutile: 1:0 (circles), 3:1  (boxes), 1:1  (diamonds), 1:3  (triangles up) and 0:1 (triangles down). (b) $R/R_0$ for the sample with 1:3 ratio of n-anatase:p-rutile under the influence of $CH_4$ (open triangles up, see also a.) and $CO$ (filled triangles up). One can see that $R/R_0$ stays (nearly) constant under $CH_4$, while it decreases considerably under $CO$. \label{bi:sav_exp}}
\end{figure}

Figures~\ref{bi:Final1}-\ref{bi:Final3} show the results of the corresponding simulations and one can see that all figures ressemble strongly the experimental data from Fig.~\ref{bi:sav_exp}(a). In the different subfigures (a)-(f), different pairs of reaction factors $(\alpha_n,\alpha_p)$ have been used that simulate the reactions to different gases. 
The np-bonds are now always considered conducting. 
The results of Figs.~\ref{bi:Final1} and \ref{bi:Final2} are attained with an occupation probability of $p=0.8$ of the sites and the material parameters of Fig.~\ref{bi:prepa}(a) and Fig.~\ref{bi:prepa}(c), respectively, while in Figs.~\ref{bi:Final3} we have $p=1$ and the parameters of Fig.~\ref{bi:prepa}(c).
To compare the reactions of the same system to different gases, we must compare curves of the same symbols (e.g. black circles for pure n-system) in the subfigures (a) to (f). 
We find selective behavior for the (blue) curves with $p_n=0.25$ and the middle (green) curves with $p_n=0.5$ that are sometimes increasing and sometimes decreasing, depending on the gas. 
The qualitative agreement with the experimental curves of \cite{Savage} is very good:  The curves of systems with a contribution of $75\,\%$ or more n-grains (two bottom curves of Figs.~\ref{bi:Final1}-\ref{bi:Final3}) always decrease with $N_r$, while curves of systems with only p-grains always increase. Between both cases, decrease or increase is possible, with a certain preference for decrease. 
The n-grains where the insulating shell is able to block the conducting channel between neighboring grains are probably more important to the specific shape of the curves than the p-grains that are always conducting. This different role of the two grain types is also reflected in two more details of the results:
First, even if $\lambda_p(N_r)$ behaves quite differently in Figs.~\ref{bi:Final1} and \ref{bi:Final2} (see insets, red stars), the curves of $R(N_r)/R(0)$ look very similar. 
This probably shows that the n-grains that are always identical have the higher influence on the results. A similar conclusion can be drawn when comparing the curves of the top rows (a-c), where the values of $\alpha_n$ are changed (and $\alpha_p$ kept constant) to the curves of the bottom row (d,e,f) where $\alpha_n$ is kept constant. In the 1st case, the curves are more volatile than in the 2nd case, which probably comes also from the higher influence of the n-grains on the results.

\unitlength 1.85mm
\vspace*{0mm}
\begin{figure}
\begin{picture}(40,30)
\def\epsfsize#1#2{0.3#1}
\put(0,1){\epsfbox{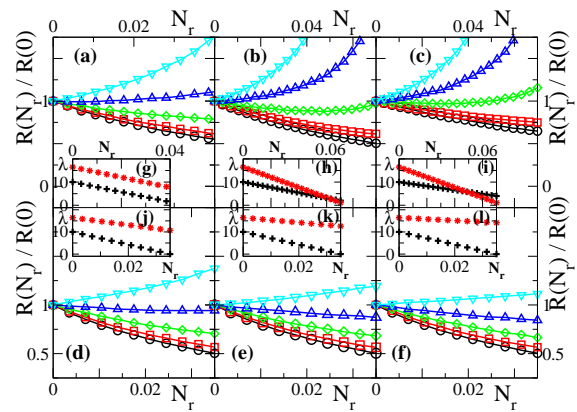}}
\end{picture}
\caption[]{\small [Color online] $R(N_r)/R(0)$ is plotted versus $N_r$ for different reaction factors (a) $\alpha_n=0.75$, $\alpha_p=1$, (b) $\alpha_n=0.5$, $\alpha_p=1$, (c) $\alpha_n=0.3$, $\alpha_p=1$, (d) $\alpha_n=1$, $\alpha_p=0.75$, (e) $\alpha_n=1$, $\alpha_p=0.5$, (f) $\alpha_n=1$, $\alpha_p=0.3$. All other parameters are the same as in Fig.~\ref{bi:prepa}(a), as well as the meaning of the different colors, symbols and of the insets (g-l), i.e. the data represent the material of Fig.~\ref{bi:prepa}(a) under the influence of different gases. 
\label{bi:Final1}}
\end{figure}

\unitlength 1.85mm
\vspace*{0mm}
\begin{figure}
\begin{picture}(40,30)
\def\epsfsize#1#2{0.3#1}
\put(0,1){\epsfbox{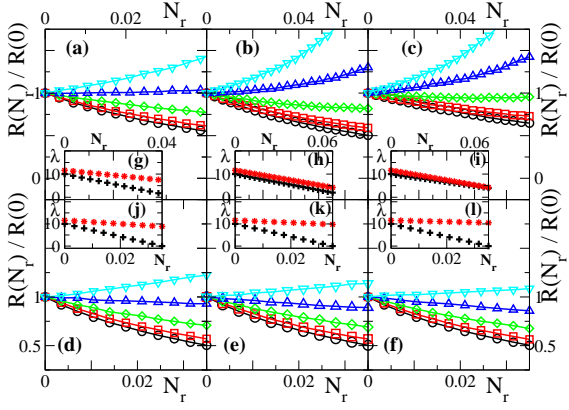}}
\end{picture}
\caption[]{\small [Color online] $R(N_r)/R(0)$ is plotted versus $N_r$ for different reaction factors (a) $\alpha_n=0.75$, $\alpha_p=1$, (b) $\alpha_n=0.5$, $\alpha_p=1$, (c) $\alpha_n=0.3$, $\alpha_p=1$, (d) $\alpha_n=1$, $\alpha_p=0.75$, (e) $\alpha_n=1$, $\alpha_p=0.5$, (f) $\alpha_n=1$, $\alpha_p=0.3$. All other parameters are the same as in Fig.~\ref{bi:prepa}(c), as well as the meaning of the different colors, symbols and of the insets (g-l).
\label{bi:Final2}}
\end{figure}

\unitlength 1.85mm
\vspace*{0mm}
\begin{figure}
\begin{picture}(40,30)
\def\epsfsize#1#2{0.3#1}
\put(0,1){\epsfbox{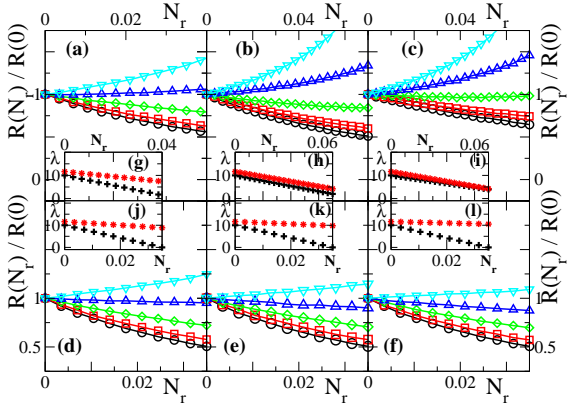}}
\end{picture}
\caption[]{\small [Color online] $R(N_r)/R(0)$ is plotted versus $N_r$ for the occupation probability $p=1$. All other parameters and the reaction factors are the same as in Fig.~\ref{bi:Final2}. 
\label{bi:Final3}}
\end{figure}

\unitlength 1.85mm
\vspace*{0mm}
\begin{figure}
\begin{picture}(40,13)
\def\epsfsize#1#2{0.3#1}
\put(0,1){\epsfbox{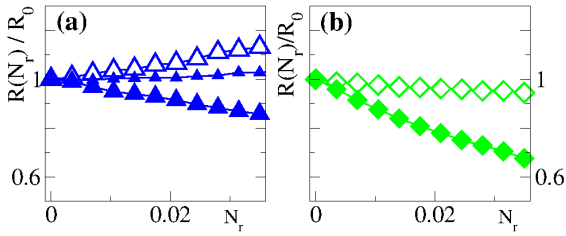}}
\end{picture}
\caption[]{\small [Color online] Some numerical data from Fig.~\ref{bi:Final2} are shown for comparison with the experintal data from Fig.~\ref{bi:sav_exp}. The same system (same value of $p_n$ may show increase, decrease or (nearly) constant behavior of $R$ with $N_r$. From top to bottom: (a) Curves for $p_n=0.25$ from Fig.~\ref{bi:Final2}(c - open triangles up), (a - small tringles up) and (f - big filled triangles up).
(b) Curves for $p_n=0.5$ from Fig.~\ref{bi:Final2}(c) and (f).
\label{bi:last}}
\end{figure}

In real systems of sintered grains, the geometrical distribution of the grains may be more correlated than in this work, where the particles have been distributed randomly over the lattices. Therefore, Fig.\ref{bi:Final3}, where all sites are occupied, i.e. $p=1$ has been included to estimate the influence of the specific geometrical structure of the systems, 
While the systems of Figs.~\ref{bi:Final1} and \ref{bi:Final2} have voids, the results of Fig.~\ref{bi:Final3} have been gained on compact systems. 
Even if $R(0)$ of both figures differs by a factor of roughly $2$, the values of $R(N_r)/R(0)$ are practically identical, indicating that also the geometrical structure of the system is not important for the considered question.

To conclude the basic finding of this paper, I present the results in Fig.~\ref{bi:last} in a similar way as in Fig.~\ref{bi:sav_exp}, i.e. I show one specific system in reaction to different gases (reaction factors). Fig.~\ref{bi:last}(a) shows the system of Fig.~9 with $p_n=0.25$ (blue triangles up) and $(\alpha_n,\alpha_p)$ equal to $(0.3,1)$, $(0.75,1)$ and $(1,0.3)$ and we see that the curve may increase or decrease, depending on the gas.
Fig.~\ref{bi:last}(b) shows the system of Fig.~9 with $p_n=0.5$ (green diamonds) for the two values of $(\alpha_n,\alpha_p)$ equal to $(0.3,1)$ and $(1,0.3)$. One can see that this figure is very similar to Fig.~\ref{bi:sav_exp}(b), i.e. to the experimental curves.

\subsection{V. Summary, Conclusions and Outlook}

In summary, I have developed a simple model to describe the resistance change of nanocrystalline composite systems of sintered n- and p-grains in response to different reducing gases. The systems are mapped onto discrete lattices, where the different grains are simulated by overlapping spheres and the bonds between them are estimated from the conducting overlap (channel). 

This work concentrates on the question, whether percolation effects are able to explain the experimentally observed selective behavior towards different gases. It turned out that despite several simplifications, the overall agreement with former experimental results is very good and the percolation model can explain in which way a system can distinguish between different gases, namely by the development of percolating pathways of n- and p-grains that react in different ways to the gas. The observed selective behavior of the systems can be attributed to the different responses of the space charge layers (shells) of the n- and the p-grains that can in principle be adjusted by the proper choice of the semiconducting materials and by an appropriate doping.  
In this context, it is of special interest that the np-bonds are necessary to reproduce the experimental curves. As the np-bonds provide the link between n- and p-paths, the neglection of them leads to a decoupling of both paths, leading to a less selective picture. 

Clearly, several simplifications have been applied in order to keep the numerical effort feasable: First, the gas concentration $N_r$ used in this work is a surface concentration, while in the experimental curve of \cite{Savage}, volume gas densities have been applied. The surface gas density normally depends on the volume density in a non-linear way, which influences the shape of the curves. Second, the systems have been discretized on a lattice, a procedure that is under discussion since long and it is not clear which kind of ''mixing rules'' are best suited for the present problem. The discretization, therefore, is a further simplification with the main aim to keep the parameters simple and clear. For the same reason, also the systems have been standardized: instead of the different experimental shapes and sizes for the n- and the p-grains, all grains have been assumed as spherical with the same medium size of n- and p-grains. Conductancies over next-nearest neighbors have been neglected, while in reality a multitude of paths between adjacent grains (including paths over edges and corners) may come into play. The conductances over the bonds have been estimated in a simplified way, without including e.g. different mobilities of electrons and holes.
 
Therefore, I have estimated the influence of these various factors on the global qualitative behavior by testing a lot of different values for particle sizes, occupation probabilities, width of the disorder and others. As shown in the figures, these differences are only important for the value of $R(0)$, but play nearly no role for the change of $R$ with $N_r$. Therefore, the basic challenge to find out by which mechanisms percolation effects can lead to a selective behavior of a gas sensor composed of n- and p-grains has been answered by introducing the appropriate reaction factors $\alpha_n$ and $\alpha_p$ that describe the number of gas molecules that react with the adsorbed oxygene. This mechanism is quite simple, the results are extremely stable against all kind of parameter changes other than either the strengths of the np-bonds or $\alpha_n$ and $\alpha_p$ themselves. Future work on more realistic continuous systems is of course highly desirable in order to optimize the performance of the sensors.

\section{Acknowledgements}
I gratefully acknowledge financial support from the Deutsche 
Forschungsgemeinschaft (project RU 854/2-1) and valuable discussions with Claus-Dieter Kohl, J\"org Hennemann and Tilman Sauerwald. Claus-Dieter Kohl has given many insightful advices on the physics of surfaces and of gas sensing.

\end{document}